\documentclass[11pt,twoside]{article}
\usepackage{asp2010}
\usepackage{url}
\usepackage{natbib}

\newcommand{\cmK}{cm$^{-3}$\,K}

\resetcounters

\begin{document}

\title{IBEX, SWCX and a Consistent Model for the Local ISM}
\author{Jonathan D.\ Slavin\affil{Harvard-Smithsonian Center for
Astrophysics, MS 83, 60 Garden Street, Cambridge, MA 02138, U.S.A.}}

\begin{abstract}
The Local Interstellar Medium (LISM) makes its presence felt in the
heliosphere in a number of ways including inflowing neutral atoms and dust and
shaping of the heliosphere via its ram pressure and magnetic field.  Modelers
of the heliosphere need to know the ISM density and magnetic field as boundary
conditions while ISM modelers would like to use the data and models of the
heliosphere to constrain the nature of the LISM.  An important data set on the
LISM is the diffuse soft X-ray background (SXRB), which is thought to
originate in hot gas that surrounds the local interstellar cloud (LIC) in
which the heliosphere resides.  However, in the past decade or so it has
become clear that there is a significant X-ray foreground due to emission
within the heliosphere generated when solar wind ions charge exchange with
inflowing neutrals.  The existence of this SWCX emission complicates the
interpretation of the SXRB.  We discuss how data from IBEX and models for
the Ribbon in particular provide the possibility of tying together heliosphere
models with models for the LISM, providing a consistent picture for the
pressure in the LISM, the ionization in the LIC and the size and shape of the
heliosphere.
\end{abstract}


\section{Introduction}
The Local Insterstellar Medium (LISM) is dominated by the large, $\sim 100$ pc
radius, very low density region known as the Local Bubble or Local Cavity.
This region is irregularly shaped and its extent has been mapped out via a few
different methods such as dust extinction and Na\,\textsc{i} and
Ca\,\textsc{ii} absorption lines \citep{Welsh_etal_2010} which probe the
distance at which substantial extinction or absorption begin.  Within the
bubble are several warm, $T \sim 7000$ K, partially ionized clouds
\citep{Redfield+Linsky_2008,Frisch_etal_2011}, known as the Complex of Local
Interstellar Clouds (CLIC), including the Local Interstellar Cloud (LIC) that
surrounds the heliosphere.


\section{The Diffuse Soft X-ray Background}
\begin{figure}[ht!]
  \centering
  \includegraphics[height=.2\textheight]{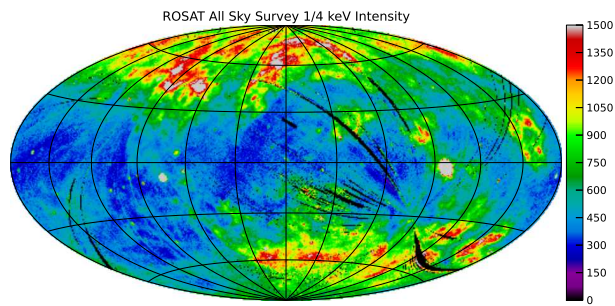}
  \caption{ROSAT all sky map in the R1 and R2 band which peaks in sensitivity
at around 1/4 keV.  The high intensity of the background at high latitudes
led to models that put the emission in the halo, but that model would predict
zero emission in the Galactic plane since the X-rays would all be
absorbed.\label{fig:SXRB}}
\end{figure}

Diffuse soft ($< 1$ keV) X-rays were first detected more than forty years ago
\citep[and references therein]{Williamson_etal_1974}. Though the brightening
of the emission toward the poles suggested emission from a hot Galactic halo
(see figure \ref{fig:SXRB}), it was soon determined that the simplest model
for the source, radiation from a hot Galactic halo absorbed by Galactic
H\,\textsc{i} (and other accompanying elements), was not adequate.  The
anti-correlation between the H\,\textsc{i} column density and the soft X-ray
flux was not strong enough to support the absorbed halo model.  Other data,
such as the lack of variation in the ratio of flux in different energy bands
even as the total intensity varied strongly, suggested that the X-rays are
generated nearby with little absorption \citep{Snowden_etal_1990}.  For the
lowest energies ($\sim 100 - 284$ eV) especially, the X-ray emission must be
from nearby because the photoelectric cross section of H and He at those
energies is so large that a small column density of neutral gas causes
substantial absorption.  However, an only slightly more complicated
model did fit the data well, the so-called displacement model. In this model
it is assumed that all the emission is generated by a uniform emissivity,
single temperature hot plasma contained in the Local Bubble.  The emission
variations are then explained as variations in the emission path length.
Combining this model with estimates of the distance to the edge of the bubble
can then yield an estimate for the thermal pressure in the Local Bubble.  This
model did a fairly good job of explaining the emission morphology, but
predicted a relatively high pressure for the Local Bubble of $P/k_\mathrm{B}
\approx 1 - 2 \times 10^4$ \cmK. 

This determination leads to problem no.~1 for Local Hot Bubble (LHB) model,
the excess pressure problem.  The high thermal pressure predicted by the LHB
model does not match the thermal pressure determined for the CLIC and in
particular for the LIC. Absorption line and \emph{in situ} neutral density and
temperature data on the LIC gives much lower value, thermal pressure $\sim
2000 - 3000$ cm$^{-3}$ K. Magnetic pressure could help balance out this
discrepancy if the magnetic pressure in the LIC could make up for the
difference in thermal pressures. However, as we discuss below, the evidence
suggests that the field is not large enough to make up for a difference of
$P/k_\mathrm{k} \ge 7000$ \cmK, which would require $B \ge 5\,\mu$G.
Heliosphere models give us our best handle on the field strength in the LIC
and most models that are consistent with the size of the solar wind
termination shock as determined by the Voyager crossings have a magnetic field
strength lower than needed to help balance the derived LHB thermal pressure.
We note that without pressure support, cloud would be crushed on dynamical
time scale $\sim L/c_s \sim 10^5$ yr, where $L$ is the scale of the cloud and
$c_s$ is the signal (sound or magnetosonic) speed. In addition, the motion of
the LIC relative to the Sun is determined by a large number of absorption line
measurements across a large fraction of the sky.  These velocity components
are consistent with a single LIC velocity vector to high accuracy
\citep{Frisch_etal_2011} and there are no indications of significant dynamical
motions occurring in the LIC. 

Problem no.~2 for the LHB model is that there is another potential nearby
source of diffuse soft X-rays. Charge exchange between inflowing interstellar
neutrals and highly ionized solar wind ions (e.g.\ O$^{+6}$, O$^{+7}$) leads
to X-ray emission (SWCX) that’s hard to distinguish from hot gas emission at
the low spectral resolution of most of the diffuse soft X-ray background
observations.  Some models for the emission find up to 100\% of the
diffuse emission observed at low energies by ROSAT in the galactic plane can
be explained as SWCX. \citet{Lallement_2004} showed that the ROSAT intensity
distribution could be explained by a combination SWCX and LHB emission with a
large fraction of the emission coming from SWCX.

SWCX modeling is somewhat complex, however, and the results depend on several
factors.  Among the requirements of the models are: the solar wind
characteristics -- speed, density, ionization, heavy element abundances, 
characteristics of the interstellar inflow -- speed, density of H$^0$, He$^0$,
and atomic data -- cross sections for capture into different
energy levels, branching ratios for cascades. The SWCX emission is variable
both temporally and spatially and is different in the slow and fast solar wind
\citep{Koutroumpa_etal_2009,Koutroumpa_etal_2011}. For this reason and because
some of the atomic data is uncertain, constraining the emission in the LHB
using SWCX alone is not currently feasible.

The soft X-ray background observations over much of the sky were done with low
spectral and spatial resolution detectors such as the proportional
counter of ROSAT and those of the Wisconsin sounding rocket observations.
Higher spectral resolution X-ray observations of diffuse emission have been
done, though over more limited regions of the sky. The Chandra X-ray
Observatory has excellent spatial resolution, $\sim 2$ arcsec, and better
spectral resolution than proportional counters.  The high spatial resolution 
allows for good point source removal, but unfortunately Chandra's ACIS
detector is not sensitive at energies below $\sim 0.4$ keV.  Nevertheless we
have used the long observations taken in the Chandra Deep Field South to try
to extract the diffuse emission in the LHB and SWCX emission simultaneously
\citep{Slavin_etal_2013}.   Our fits to the diffuse emission require
significant hot gas emission.

A new analysis of old data also provides insights into the LHB emission.
The Diffuse X‐ray Spectrometer (DXS) experiment was flown in the 90’s
\citep[see][for details]{Sanders_etal_2001}.  In comparison to the
proportional counter instruments it had good spectral resolution though poor
spatial resolution.  The results were quite interesting, showing a spectrum
with several different lines but no acceptable fits were found to the spectrum
using standard models.  Recently a group at the CfA revisited 
fitting the data and have new modeling results (R.\ K.\ Smith et al.\ 
submitted) using a combination of slow and fast solar wind, SWCX and LHB
emission that finally give good fits.  Using the model results and the
estimated distance to the edge of the LHB for the direction observed they
derive a thermal pressure in the Local Bubble of $P/k_\mathrm{k} = 5800$
\cmK.  This value certainly has significant uncertainty attached to it, but
nonetheless it is interesting as it is much more in line with our expectations
given the pressure in the LIC.

\section{The Interaction of LHB Hot Gas and the LIC}
Fairly straightforward analysis based on absorption line data toward nearby
stars indicates that the LIC is partially ionized, $X(\mathrm{H}^+) \sim
30$\%, $X(\mathrm{He}^+) \sim 40$\%, far more so than would be expected from
collisional ionization at the derived temperature of $\sim 6500$ K.  Thus an 
ionization source is required. There are few main sequence hot stars inside
the Local Bubble and a census of them and hot white dwarfs carried out by the
Extreme Ultraviolet Explorer (EUVE) satellite found that they do not produce
enough ionizing photons to account for the ionization of the LIC
\citep{Vallerga_1998}.  The only other identified sources of ionizing radiation
are the hot gas of the LHB and the interaction region at the boundary of the
LIC and the hot gas \citep{Slavin_1989}.

Thermal conduction between the surrounding hot gas ($T \sim 10^6$ K) and the
warm LIC leads to evaporation of the cloud.  Intermediate temperature gas ($T
\sim 10^5$ K) in the boundary radiates strongly in the EUV and is therefore a
good ionization source \citep{Slavin_1989,Slavin+Frisch_2008}. The magnetic
field support for the cloud drops off in the boundary as the density drops and
is insignificant in hot gas, so the model requires that the thermal $+$
magnetic pressure of the cloud equals the thermal pressure of the LHB to
maintain pressure equilibrium.

\begin{figure}[ht!]
  \centering
  \includegraphics[height=.3\textheight]{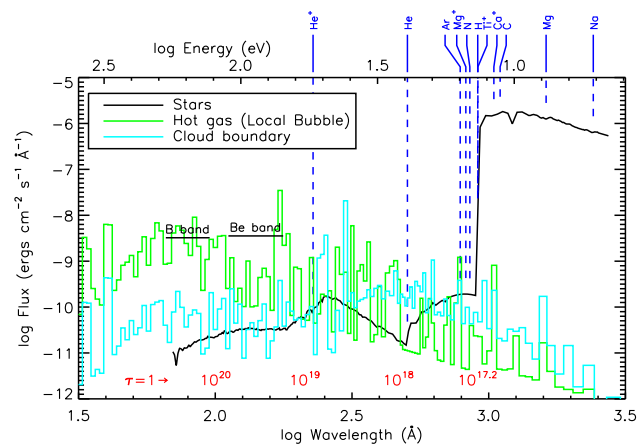}
  \caption{Modeled radiation field incident on the LIC.  The black line shows
the flux from stars including nearby hot stars and, for wavelengths above
912\AA, the flux from many cooler stars.  The green line is the modeled
emission from the bulk of the volume of the LHB and depends on the thermal
pressure in the bubble.  The blue line is for emission that comes from the
boundary region between the warm cloud and the surrounding hot gas.
Ionization thresholds for several ions are shown at top and the wavelength at
which optical depth unity is achieved for several H\,\textsc{i} column density
values is shown in red.  [From \citet{Slavin+Frisch_2008}.]
\label{fig:spectrum}}
\end{figure}

As mentioned above, the SWCX contribution to the soft X-ray background reduces
the amount that can be attributed to the hot gas emission.  In this way the
SWCX contribution to soft X-ray background lowers the pressure demand in the
LHB -- in other words problem no.~1 for the LHB can be solved by problem
no.~2.  But one can go too far with this.  We still need the pressure of the
hot gas to confine the cloud -- no other gas phase is viable without incurring
excessive energy supply or emission problems.  This leads to the question: Can
the lower pressure hot gas still create enough ionizing flux to ionize the
LIC? 

Our models of the photoionization of the LIC
\citep{Slavin+Frisch_2008} include the radiation field from nearby hot stars,
emission from the LHB and emission from the boundary region between the hot
gas and the LIC where the gas is evaporating off the cloud (see figure
\ref{fig:spectrum}).  It was assumed in the models that all of the soft X-ray
background came from the LHB and thus all those photons were available to
ionize the LIC.  SWCX emission is moderately bright relative to the LHB
emission at the Earth because it is generated so close to us, but it is
insignificant for the ionization of the LIC.  Thus a lower fraction of the
soft X-ray background that is from the LHB, the less flux of soft X-rays that
are available for ionizing the LIC.  However, the soft X-rays are fairly
inefficient for the ionization of H and He, so it is the associated
diffuse EUV radiation, which has not been directly observed, that is
responsible for the photoionization.  In the model most of that emission
originates in the cloud boundary region.

\begin{table}[ht!]
\caption{Model Results with Reduced B band Emission}
\small
\begin{tabular}{cccccccccc}
\tableline
&  & \multicolumn{8}{c}{Percentage of B band Emission from
the Local Bubble} \\
\cline{3-10}
\multicolumn{2}{c}{Model Parameters} & 100\% & 50\% &
100\% & 50\% & 100\% & 50\% & 100\% & 50\% \\ 
\cline{1-2} $N_\mathrm{HI}/10^{17}$ & $\log T_\mathrm{hot}$(K) &
\multicolumn{2}{c}{$B_\mathrm{LIC}$($\mu$G)} & 
\multicolumn{2}{c}{$n$(H$^+$)\,(cm$^{-3}$)} &
\multicolumn{2}{c}{$n$(H$^0$)\,(cm$^{-3}$)} & 
\multicolumn{2}{c}{$P_\mathrm{LB}$(cm$^{-3}\,$K)} \\
\tableline
 3 & 5.9 & 4.63 & 5.60 & 0.077 & 0.082 & 0.186 & 0.177 & 8680 & 11400 \\
 4 & 5.9 & 2.52 & 3.74 & 0.055 & 0.058 & 0.191 & 0.189 & 4080 & 6280 \\
 3 & 6.0 & 4.06 & 5.03 & 0.070 & 0.073 & 0.197 & 0.194 & 7170 & 9750 \\
 4 & 6.0 & 2.05 & 3.47 & 0.051 & 0.054 & 0.193 & 0.192 & 3300 & 5600 \\
 3 & 6.1 & 3.51 & 4.65 & 0.065 & 0.070 & 0.204 & 0.201 & 5710 & 8440 \\
 4 & 6.1 & 0.05 & 3.22 & 0.050 & 0.052 & 0.197 & 0.193 & 2150 & 4920 \\
\tableline
\label{tab:modparams}
\end{tabular}
\normalsize
\end{table}

To explore the effects of having a substantial amount of the soft X-ray
emission come from SWCX instead of the LHB we have recalculated our
photoionization models with 50\% of the Wisconsin B band (100 - 284 eV)
emission missing.  By necessity the pressure in the hot gas is equal to that
in cloud and since we control the cloud parameters the hot gas pressures find
their own level.  The thermal pressure in the boundary sets the emissivity
there and other model parameters, e.g.\ elemental abundances, are set by
matching observed column densities \citep[see][for
details]{Slavin+Frisch_2008}.  Less emission from the LHB requires more from
cloud boundary and thus we find that our new models need a somewhat
higher magnetic fields than our old models.  Results for these models are
listed in table \ref{tab:modparams}. We find that a subset of the
models are consistent with LHB thermal pressure of $\sim 5800$ \cmK\ as
determined for the DXS data, which given the LIC thermal pressure implies
$B_\mathrm{LIC} = 3.4 - 3.6$ $\mu$G. 
\begin{table}[ht!]
\centering
\caption{Heliosphere model parameters}
\begin{tabular}{cll}
\tableline\
$B_\mathrm{LIC}$ ($\mu$G) & $n(\mathrm{H}^+)$ (cm$^{-3}$) & 
$n(\mathrm{H}^0)$ (cm$^{-3}$) \\
\tableline
2 & 0.13 & 0.22 \\
3 & 0.095 & 0.195 \\
4 & 0.048 & 0.16 \\
\tableline\
\label{tab:JH}
\end{tabular}
\end{table}

The magnetic field is best constrained by heliosphere models and
data. \cite{Schwadron_etal_2011} used IBEX ENA and other data and a
semi-analytic model and estimated the interstellar magnetic field to be $\sim
3\,\mu$G.   The IBEX Ribbon location and strength depends on $B_\mathrm{LIC}$,
the magnetic field in the LIC, and the proton and neutral H density and
velocity at the heliosphere. Recent results by Heerikuisen (in preparation)
finds good fits to the heliosphere shape with the parameters in table
\ref{tab:JH}. The 4 $\mu$G case leads to a ribbon that does not fit the data
well.  \citet{Zank_etal_2013} also find a $B_\mathrm{LIC} \sim 3$ $\mu$G fits
observations of the H wall and 4 $\mu$G does not. 

\section{Conclusions}
With the combination of the soft X-ray diffuse background observations, data
from IBEX, especially the ribbon, and absorption line data toward nearby stars
the constraints on the nature of the LIC are becoming ever tighter.  The links
between the photoionization of the LIC and the heliosphere are also getting
stronger because, if the ionizing radiation field includes a substantial
contribution from the emission generated at the cloud edge, then the
ionization of the cloud depends on the strength of the interstellar magnetic
field as does the shape of the heliosphere.  Charge exchange reactions between
inflowing interstellar neutrals and solar wind ions play important roles in
shaping the heliosphere and in generating emission that makes up part of the
soft X-ray diffuse background.  The new interpretation of the DXS data by
R.~Smith et al.\ lends additional credence to the idea that the background is
generated both by SWCX and by hot gas in the Local Bubble and that there is a
consistent pressure for the hot gas and the LIC, which is partially supported
by the magnetic field.  This pressure is lower than previous estimates that
did not include the SWCX emission, but hot gas is still required to fill the
Local Bubble.  Further modeling of the SWCX emission, the heliosphere and
photoionization of the LIC needs to be pursued to fully realize the goal of a
consistent model for the LISM, heliosphere and LHB, but we are closer than
ever to achieving that goal.

\acknowledgements{I thank my collaborators Priscilla Frisch, Nicholas
Pogorelov, Jacob Heerikhuisen, Hans-Reinhard M\"uller, Bill Reach and Gary
Zank who have worked with me on problems related the interaction of the LIC
with the heliosphere and the conference organizers for inviting me to give
this talk.  This work was supported by IBEX project which is funded by NASA as
part of the Explorer Program.}


\end{document}